# Blockchain based secure energy marketplace scheme to motivate peer to peer microgrids

Muhammad Awais[1], Qamar Abbas[2], Shehbaz Tariq[1], Sayyaf Haider Warraich[2]
[1]Electronics and Information Convergence Engineering, Kyung Hee University, Seoul, South Korea
[2]National University of Sciences and Technology, Islamabad, Pakistan

**ABSTRACT**

In the past few years, the trend of microgrids has been increasing very fast to reduce peak-hour costs. However, in these systems, third parties are still involved in selling surplus energy. This raises the cost of energy, and such systems have numerous operational and security barriers. These issues can be solved by the decentralized distributed system of microgrids, where a consumer can locally sell their surplus energy to another consumer. To deploy such a system, one must consider security barriers for the transaction of energy. This paper proposes a solution to these problems by devising a scheme as a marketplace where users interact with each other to buy and sell energy at better rates and get energy-generating resources on lease so that users do not have to worry about capital investment. An agreement between the owner of resources and the consumer is recorded on blockchain-based smart contracts. In this paper, a survey of well-known decentralized energy solutions is conducted. This paper also proposes an extra layer of security to leverage a shielded execution environment so that information about energy generated, utilized, and shared cannot be changed by consumers and third parties even if the system is compromised.

*Corresponding Author:*

Muhammad Awais
Electronics and Information Convergence Engineering, Kyung Hee University
26 Kyungheedae-ro, Dongdaemun-gu, Seoul, Yongin, South Korea
Email: mawais@khu.ac.kr

## 1. INTRODUCTION AND MOTIVATION

Microgrids are distributed energy sources with clearly defined boundaries concerning grids, which were introduced to fulfil local energy demands and to increase the efficiency and flexibility of distributed energy resources [1], [2]. From the start, our energy distribution systems were designed on a large scale by considering the problems of macro-scale distributions, e.g., demand, abruptions. However, with time, use of microgrids increased, which allowed the consumers to generate their energy to decrease peak hour load charges while increasing renewable energy resources [3]. Currently, the use of low-voltage energy services like electric vehicles, smart appliances, photovoltaic panels, and battery-based storage systems is being increased, and the optimised use of microgrids can improve reliability, reduce cost, and integrate more efficiently with the grid; that is why regulation and some policies have been introduced [4]. Prosumers who have surplus energy require extra energy to sell, but current systems require third parties to be involved in selling and buying such energy. The involvement of third parties has increased the cost of this energy, which is why consumers have to pay more and prosumers get less by selling their extra energy. In conventional systems, consumers cannot identify the energy source they consume, nor can they buy it from specific prosumers. Integrating such new features is very challenging in conventional systems. Due to this, both parties rely on the third party, which consumers and prosumers do not desire.

Since the Bitcoin white paper [5] publication by an anonymous person named Satoshi Nakamoto, a new technology called blockchain has been introduced whose purpose is mainly to create a decentralised currency to mitigate monopolies by banks. The blockchain ledger uses a decentralised peer-to-peer mechanism in which every node is connected to one or multiple other nodes in a network. Whenever a new block is created or a transaction is performed, then randomly selected nodes perform cryptographic calculations to validate the transaction, and whenever a new transaction is validated, then nodes broadcast this information to other connected nodes. Different consensus protocols can solve and validate such a canonical ledger of blockchain, and there is a possibility of multiple segmentations of the ledger, but only then the longest chain is used as an official ledger which is trusted and shared by the majority of nodes [6], [7].

Due to the considerable promise of decentralisation and security, blockchain technology is now being applied in many other countless applications, e.g., supply-chain management, health care industries, cryptocurrencies, decentralised cloud data, notary services, the music industry, voting systems [8], and many other applications in which automatic implementation according to a set of rules is required using smart contracts. Similarly, blockchain technology can completely decentralise microgrids where a consumer can sell surplus energy to consumers without any third-party involvement. Such a system can result in high profits for prosumers as they do not have to pay third parties to sell their extra energy, and consumers can buy energy at lower rates as they do not have to pay third parties for their services. Consumers in our proposed system can choose the amount of energy, source of energy generation, and microgrid from which they want to purchase energy. Pollution is one of the most serious issues confronting modern society, owing to the fact that most consumers would prefer to purchase renewable energy [9] at the lowest possible cost from prosumers. Our system scheme creates fair competition between prosumers to sell energy at a fair price and promotes the generation of more renewable energy.

Blockchain technology in microgrids can deliver more secure, reliable, and cost-efficient energy sharing. Blockchain technology can create new energy trading systems with improved pricing, transparency, and speed [10]. Most people don't trust blockchain, despite its security. We propose shielding hardware security for the market. Users can verify the server's running code using a PKI certificate. Shielded execution is new and depends on the hardware. Like Intel-SGX [11], it's utilized on Intel-based servers and trusted-zone [12] on ARM-based servers. Intelligent meters with protected execution can add security to blockchain. Using the technique can dynamically interconnect microgrids with the grid, enabling intelligent grids and smart cities. Blockchain-based smart microgrids are effective in remote places with frequent line breakage, large line losses, and high capital costs. In impoverished countries, such system installation is easier than in rich ones, and the community can benefit from renewable and cheap energy. This paper's primary contributions are: i) we proposed an energy marketplace in which flexible features can be introduced with mutual agreement of consumers, prosumers, and resource owners, which provides a fair trade of energy and an investment opportunity; ii) we surveyed prominent decentralized companies and compared their focus and scope of business; and iii) to enhance the security of the proposed scheme, we leveraged shielded execution-based marketplace servers and intelliblocks. The paper is organized as follows: section 2 addresses decentralized energy literature, section 3 compares section 2 solutions, section 4 outlines the whole system, section 5 ends the paper.

## 2. LITERATURE REVIEW

The decentralized nature of blockchain is the best option for decentralized microgrid systems. Many projects are using blockchain in microgrids, but most of the projects are under the development phase. Currently, it is hard to setup a new system where already excellent distribution systems are installed. However, in underdeveloped countries, it is easier to deploy blockchain based smart microgrids as new energy sources are in demand. Blockchain technology based decentralized microgrids could promote renewable energy resources by using smart contracts so that the energy source is verified, and the prosumer cannot change it later. This gives the consumer to select renewable energy sources so that such energy generation and selling is more environmentally friendly. Some exiting solutions are discussed in the following subsections.

### 2.1. SolarCoin

SolarCoin aims to promote renewable solar energy culture around the globe by encouraging consumers to invest in solar energy. Due to long payback time, prosumers invest less in solar energy. However, the company promote solar energy by giving prosumer one SolarCoin token for generating 1 MW energy from renewable solar source for which certified metered are also installed by the company [13]. SolarCoin company has a token system based on blockchain for giving rewards and buying energy.

## 2.2. The sun exchange

This project targets underdeveloped countries where corruption is a big problem. The sun exchange provides a platform where users can buy solar panels and lease solar panels to earn passion income. Solar panels owner earns money for giving his solar panels on rent for 20 years in the form of Bitcoin or local currency as per the preference of the owner [14]. The company installs these solar panels in a suitable place for which rent is given to the owner, this way the owner does not need to worry about the maintenance of solar panels, and he can earn remotely from anywhere from the globe.

## 2.3. Electron

Electron company [15] develops an Ethereum based solution of energy alongside the currently deployed system. They provide collaborative trading comparable to peer-to-peer trading. They have many tools on open source to make their smart meter more productive and secure. Currently, electron projects include recorder to create a shared asset register for energy. Artemis for flexible energy trading and Helios, a funded project for multilateral transaction platform development to have demand-side response actions.

## 2.4. PowerLedger

Based on blockchain PowerLedger provide a clearing mechanism and trading of energy. Customers can buy or sell a surplus amount of energy from renewable energy sources within microgrids or over distribution networks in real-time. Energy could be sold directly, or if a consumer has batteries, he can store energy and sell at another time for more revenue by selling at peak hour time. Buyers can see direct sellers on the platform and can get the best price. Distribution network systems also get revenue for energy trading as infrastructure is provided by distribution networks [16].

## 2.5. LO3-energy

LO3-energy is based on software and hardware, which provides its customers to sell and buy from each other using blockchain technology based on smart contracts in a securely and automatically mechanism. LO3-energy platform is based on Ethereum cryptocurrency and blockchain smart contracts to reshape the future of energy by innovations in energy buying, selling, using, storing and generation at the local level [17]. LO3-energy has the vision to create intelligent microgrid systems based on blockchain which can provide information regarding energy production, energy consumption, peer to peer transactions, and demand response. LO3-energy allows its customers a free choice to buy from multiple renewable energy sources; a customer with surplus energy can sell the energy they produce. Communities can choose to keep energy resources local to increase decentralized energy sharing efficiency.

## 2.6. Share&Charge

Share&Charge is promoting green mobility seamlessly and smartly for the future. As with the increase in the tread of electric vehicles, Share&Charge company has made a network of charging stations for electric vehicles. This way, users with surplus energy can buy a particular module from share and charge and set their tariffs for charging [18]. Persons those required to charge their electric vehicles will have to use wallets for a transaction. Share&Charge provides an easy way to charge electric vehicles even in remote areas where seller and buyer both get benefits. Share&Charge company handles billing and storing such information in the user's wallet.

## 2.7. Grid singularity

Grid singularity has created an open-source decentralized energy data exchange platform. Like other platforms, Grid singularity does not only focus on electricity trading as energy; it also focuses on gas, heating, and water in the decentralized mechanism. Grid singularity has the software developed as an agent-based model, to optimize energy devices operations. This decentralized autonomous area agent harnesses the potential of smart grid, decentralized energy sharing and renewable energy sources by data analytics and optimization [19].

## 2.8. WePower

WePower helps customers directly connect to green energy sellers at a cheap rate with good power purchase agreements. WePower [20] enables its customers to procure and trade green energy in their platform easily. Power purchase agreement helps customers buy electricity at stable rates, which are ensured by blockchain based smart contracts that automatically execute according to rules set. WePower also provides a platform to proper monitor, review contracts and find renewable electricity producers, best fit for buyers.

## 2.9. Alliander

Alliander provides a peer-to-peer innovative energy marketplace where customers can consume, produce, and share renewable energy. Alliander has good financial support and currently working on a project to make energy affordable, more accessible, and reliable in local energy networks [21]. Initially, the company was built in 2009, but decentralized projects were started in 2017. Alliander provides energy tokens to its customers; these could be used in more energy requirement times or consumers can sell these tokens to generate revenue.

### 2.10. Energy web foundation

Energy web foundation (EWF) [22] was started as a non-profit organization by world-leading energy sector companies to use blockchain technology potential in the energy sector. EWF is one of the top enterprise-grade platforms which use blockchain in energy regulation, operations and solving market energy need problems. EWF is built to handle a large number of transactions with high speed and scalability.

### 2.11. NRGcoin

NRGcoin was started as an industry-academia linked project, which was latterly scaled up to an industrial context [23]. NRGcoin uses smart contracts to promote green energy by providing benefits in smart grid for energy producers, consumers, government, and energy distribution network providers. In NRGcoin scheme, energy producers can sell extra energy to earn NRGcoins, and incentives to produce renewable energy, and consumers get cheap energy using NRGcoin, the government gets tax and DSO gets a margin by providing the infrastructure which encourages customers to use the NRGcoin platform [24].

## 3. COMPARISON OF PEER TO PEER ENERGY SHARING INDUSTRIES

In Table 1, a brief comparison of early mentioned industrial power sector projects based on blockchain is provided. In some previous papers like [25]–[27], comparisons having some similarities are provided, and we have also considered their provided information for comparing these industrial projects. There are a lot of non-industrial projects that also exist for example [28]–[30] but we are not considering those in our survey. We have not considered all the industries in our comparison as too many industries are being established day by day in this market. In the time of availability of this research to readers, many other new industries may be established.

Table 1. Technical review of different blockchain based microgrid industries

| Projects | Short description | Blockchain ledger | Type of blockchain | Consensus mechanism | Open source | Foundation year |
|---|---|---|---|---|---|---|
| SolarCoin | Promoting solar energy culture by giving SolarCoin one token for generating 1-MW. | SolarCoin | Public | PoS | Yes | 2014 |
| The sun exchange | Buy and lease solar cells to earn passive income from the globe. | SolarCoin | Public | PoS | No | 2015 |
| Electron | Collaborative energy trading provides asset register records for energy. | Ethereum | Public | PoW | Yes | 2016 |
| PowerLedger | Platform to buy and sell energy at best rates, reduce peak hour time charges and provide revenue for DSO too. | Ethereum, EcoChain | Public, Private, | PoW, PoS | Yes | 2016 |
| LO3 energy | Working on decentralized future of energy to buy, sell, store and generate at a local level. Also working on intelligent data system in microgrids. | Ethereum | Public | PoW | No | 2017 |
| Share&charge | Promoting green mobility seamlessly and smartly for electric vehicles. | Ethereum | Public | PoS | Yes | 2017 |
| Grid singularity | Decentralized energy sharing open-source plat-form which also focuses on gas, heating and water sharing. Developed software for analytics and optimization of energy sharing. | Ethereum, EWF | Public | PoW, PoA | Yes | 2016 |
| WePower | Provide power purchase agreements for long term stable and good rates. Also, provide a platform for the management of trades. | Ethereum | Public | PoW | Yes | 2016 |
| Alliander | P2P energy trade marketplace to buy, sell and share energy. | Ethereum | Public | PoS | No | 2017 (build 2009) |
| Energy web foundation | Use blockchain in energy regulation, operations and solving problems. EWF is built to handle a large number of transactions. | Ethereum | Public | PoA | Yes | 2017 |
| NRGcoin | Started as an industry-academia project to sell energy in a decentralized way. | Ethereum, NRGcoin | Public | PoW, PoS, PoA | Yes | 2014 |

## 4. PROPOSED SYSTEM MODEL
### 4.1. Problem

Companies that want to provide a marketplace for buy, sell, and share energy generated will have to make sure all problems related to blockchain-based microgrids are solved related to the accuracy of data, traceability of the origin of energy, privacy in individual level of users and security of the blockchain. For this purpose, good quality smart digital meters are essential so that other customers are satisfied with this new technology. There has been much work already done for P2P energy trading, not just to make it decentralised but also to make it cheaper and easily accessible. Most of the projects promote renewable solar energy, but with a change in the environment, only one renewable energy source is not enough; therefore, one should utilise multiple resources for effective energy harvesting from available energy sources. This is also very important because in different locations different environment exists so there is a high chance that if in one location solar panels are an effective way to harvest solar energy then in other location it might be wind energy as first choice and solar panels may not work effectively. Also, in areas where there is a possibility to harvest multiple renewable energy sources than in such locations, users should be encouraged to used different renewable energy sources instead of only one. This way, if one energy source is not available, then other energy sources will not be affected. It could be used to provide more reliable, balanced, and continuous renewable energy to users. There can also be cases where there is environment is not suitable for renewable energy sources or renewable energy sources lacks the demand from consumers in those areas other energy sources must exploit and by using those sources in a shared manner can reduce the burden on one consumer and even one prosumer can earn revenue by sharing energy. Another issue can be raised when a consumer does not have the resources to generate his energy. Then in such cases, an investment is required from another person with mutual benefits. Power energy distribution worldwide works better because energy is being generated not just for one user but for average society demand, which enabled efficient utilisation of energy generation. For solving these issues, a decentralised energy sharing system is required to develop, which can be leveraged to solve consumer demand and mitigate any possible conflict between investors, prosumers, and consumers. Such a system also needs to provide security guarantees of information provided to the marketplace to increase the trust factor of users. To work on such a system, we need to give a free token to prosumers to promote renewable energy. This quantity of free tokens should be adjusted so that prosumers are encouraged to use different energy sources instead of only one energy source when available. These free tokens given to customers could be redeemed in the form of money or to buy electricity.

### 4.2. System model

In proposed system scheme, there can be multiple types of agreements between users, which they can get from public records of the centralised marketplace. Agreements can include rent cost, time of the agreement, ownership of resources after an agreement, selling of energy, between the owner of energy resources, prosumers, or consumers, even with the flexibility to add a grid or third parties in agreement. Proper agreements will be made between users for energy rates and percentage rent to be paid in our marketplace. To ensure automatically contract running blockchain based smart contracts are to be made between users on agreed terms recorded in the smart contract.

In the proposed scheme, there are mainly three types of users are present, consumers who consume electrical energy, prosumers who consume as well as sell electrical energy as they have more than their required energy by installing energy harvesting resources and owners who want to invest in buying resources and giving to users on rent and sell energy at a cheap rate to them. Whenever an owner installs energy harvesting resources to its building, then it also becomes a prosumer. In the marketplace, users can also be some other thirds parties and even government agencies. Third parties and government agencies can act as any other three types of users too. Most utilisation of grid to transfer energy from prosumer to consumer will also result in giving rent to grid provider, and this will highly depend on the country in which microgrid is installed and their policies. That is why we proposed a scheme as shown in Figure 1 with high flexibility in its agreements (smart contracts).

In an agreement, one (owner) can buy a renewable energy harvesting system and can lease it to consumers who have space to install this system. In this scheme, the owner invests in a renewable energy harvesting system to earn revenue by selling energy to the consumers. In the scheme, the energy system will be installed on-premises of consumer, and the consumer will pay for energy utilisation, and a percentage of those charges will be returned to the consumer as rent money of premises underuse. This is useful for the owner and the consumer as the owner will get revenue by selling energy, and the owner does not need to install an energy system on his premises. A consumer can benefit by purchasing energy at better rates and getting discounts due to providing his premises space for the energy system. If the energy system is generating more energy than the consumer's requirement, then by selling surplus energy to nearby other consumers, owner and consumer both can get benefit by sharing revenue on the agreed terms of smart contracts. This agreement is helpful for consumers living in remote areas where energy is not provided and consumer cannot afford to buy own renewable energy system to generate energy. This agreement can be subdivided into two types of categories of smart contracts, one in which the owner will sell energy system to the consumer in a specified period hence owner does not have to pay rent; instead, he gets a fixed amount from consumers in instalment to sell his energy harvesting system and till that time owner gets revenue for selling energy to the consumer. In another type of smart contract consumer only pays for energy used and get a discount for providing a place to install the system. In this scheme, the owner will get back his energy system from the consumer after the specified time in a smart contract, generally taken as 10-15 years for which consumers pay energy utilisation charges to the owner.

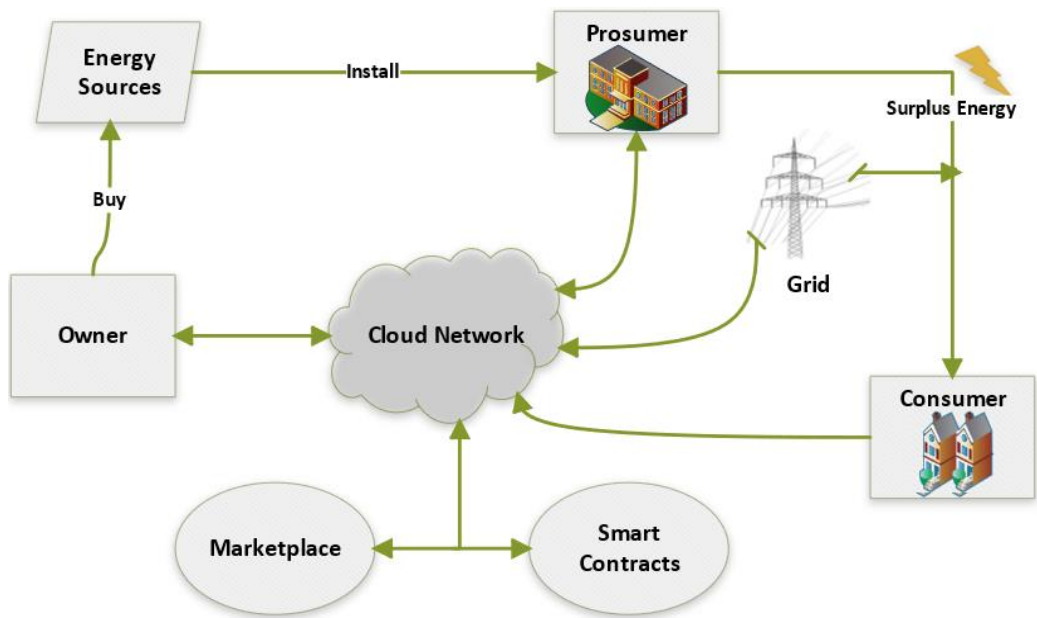

Figure 1. The architecture of the system scheme

Our marketplace will be providing information of different owners with their capital investment in energy systems and consumers interested to install such energy systems on their premises. Owners and consumers will interact with each other in our marketplace. They will set terms and conditions that will be saved in our marketplace's database and in a backend blockchain ledger to ensure smart contracts' automatic execution in a later stage, which could be used as legal proof if any party breaks set rules. In marketplace information of owners, type of energy harvesting system they own, capacity of energy system, location preferences, information of consumers, size of premises to install energy system, energy system type preference of consumers, consumer's location, and any other additional information will be provided. In the marketplace, a rating system is to be introduced to interact with more trusted users, which will encourage them to build trust in our marketplace. Our marketplace will be in synchronous mode with a blockchain ledger to achieve better security and reliability. As the energy resources will be installed in different areas, security and data integrity are very important so that consumers cannot modify any informative data related to energy harvested and utilised. For the security of such data, we provide solutions based on shielded execution and due to shielded execution-based hardware (smart meter and reading sensors), consumers will be unable to modify any data, which is a crucial requirement as a consumer can be living in a remote place and physical monitoring will not be feasible. This security feature will also secure third parties' invasion of the system in a compromised environment. Just like securing consumer side shielded execution-based hardware security will be leveraged in the marketplace's server to provide data integrity and confidentiality which can increase the trust of users in the marketplace.

## 5. CONCLUSION

The use of new blockchain technology as a smart contract of energy sharing is very useful and efficient as no third parties are involved, and a better energy sharing rate could be achieved. In the proposed model, the owners can rent their energy generating resources to consumers having a place to install, which is greatly useful for prosumers as well as for consumers to get revenue and better rates of electricity respectively, while promoting utilization of renewable energy sources. In under developed countries, it is easy to deploy decentralized schemes as a less changing requirement with current infrastructure, and if new infrastructure is required to be installed, then it can be made capable of handling decentralized energy sharing. Also, currently, as blockchain technology is still emerging and has its vulnerabilities at commercial levels, that's why customers are not yet relying on it. As it is hard to motivate people due to lack of trust in new technologies so added security by shielded execution and incentives of generating renewable energy can motivate more people to use decentralized renewable energy systems.

## BIOGRAPHIES OF AUTHORS

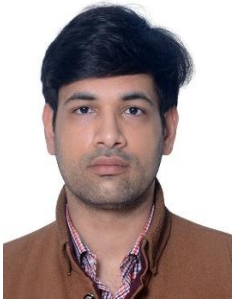
**Muhammad Awais** is currently a Ph.D student at Kyung Hee University, South Korea. His area of research is wireless communication, reconfigurable intelligent surface, and machine learning. Previously he has been part of different R&D industrial projects for four years. He has completed his MS in electrical engineering from the National University of Sciences and Technology, Pakistan, and B.Sc in electrical engineering from UET, Lahore, Pakistan in the year 2020 and 2016 respectively. He can be contacted on email at: mawais@khu.ac.kr.

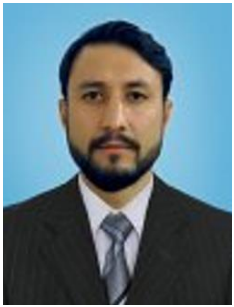
**Qamar Abbas** (bachelor of science) degree in electrical engineering in 2014 from Comsats University Islamabad Pakistan. Then completed his MSc in electrical engineering in 2017 from the University of Lahore (Islamabad Campus). He is currently a Ph.D student at the University of Sciences and Technology, Pakistan and working at Air University, Pakistan as a research associate. His research interests include age of information, markov analysis, NOMA, stochastic systems, and backscatter communication. He can be contacted on email at: qamar.abbas@uaf.edu.pk.

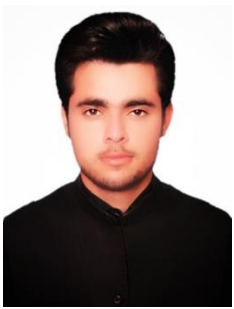
**Shehbaz Tariq** is currently an MS student at Kyung Hee University, South Korea. He has completed his B.Sc in Electrical Engineering from the University of Engineering and Technology (UET), Peshawar, Pakistan in September 2020. His research interests include quantum information science, wireless communication, and machine intelligence. He can be contacted on email at: mawais@khu.ac.kr.

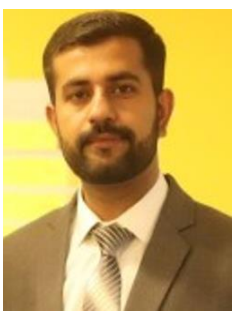
**Sayyaf Haider Warraich** is currently working as Founder and Chief Operating Officer (COO) at XpertFow Technologies since 2019. He has been associated with industrial R&D projects for the last 5 years. He has completed his MS in electrical engineering from the National University of Sciences and Technology, Pakistan in the year 2020 and his B.Sc in electrical engineering from GCU Faisalabad, Pakistan in the year 2015. His research interests include healthcare, machine intelligence, and SDN. He can be contacted on email at: sayyafhaider.msee22@students.mcs.edu.pk.